# On the computational solution of vector-density based continuum dislocation dynamics models: a comparison of two plastic distortion and stress update algorithms

Peng Lin[1], Vignesh Vivekanandan[1], Kyle Starkey[1], Benjamin Anglin[2], Clint Geller[2], Anter El-Azab[1]

[1] Purdue University, West Lafayette, IN 47907, USA

[2] Naval Nuclear Laboratory, West Mifflin, PA 15122, USA

**Abstract**

Continuum dislocation dynamics models of mesoscale plasticity consist of dislocation transport-reaction equations coupled with crystal mechanics equations. The coupling between these two sets of equations is such that dislocation transport gives rise to the evolution of plastic distortion (strain), while the evolution of the latter fixes the stress from which the dislocation velocity field is found via a mobility law. Earlier solutions of these equations employed a staggered solution scheme for the two sets of equations in which the plastic distortion was updated via time integration of its rate, as found from Orowan's law. In this work, we show that such a direct time integration scheme can suffer from accumulation of numerical errors. We introduce an alternative scheme based on field dislocation mechanics that ensures consistency between the plastic distortion and the dislocation content in the crystal. The new scheme is based on calculating the compatible and incompatible parts of the plastic distortion separately, and the incompatible part is calculated from the current dislocation density field. Stress field and dislocation transport calculations were implemented within a finite element based discretization of the governing equations, with the crystal mechanics part solved by a conventional Galerkin method and the dislocation transport equations by the least squares method. A simple test is first performed to show the accuracy of the two schemes for updating the plastic distortion, which shows that the solution method based on field dislocation mechanics is more accurate. This method then was used to simulate an austenitic steel crystal under uniaxial loading and multiple slip conditions. By considering dislocation interactions caused by junctions, a hardening rate similar to discrete dislocation dynamics simulation results was obtained. The simulations show that dislocations exhibit some self-organized structures as the strain is increased.

# 1 Introduction

The continuum dislocation dynamics approach to crystal deformation has gained a great deal of attention lately (El-Azab, 2000; Groma et al., 2003; Hochrainer, 2016; Hochrainer et al., 2014; Monavari et al., 2016; Monavari and Zaiser, 2018; Sudmanns et al., 2019; Xia and El-Azab, 2015; Yefimov et al., 2004). This approach is attractive due to the fact that one can use it to integrate the mechanics and physics of dislocations into continuum mechanics descriptions of metal deformation in much the same manner as with discrete dislocation dynamics methods (Cui et al., 2019, 2014; Devincre et al., 2008, 2006; El-Awady, 2015; Liu et al., 2017; Sills et al., 2018; Stricker et al., 2018; Stricker and Weygand, 2015). While the latter methods proved to be powerful in tackling important mesoscale plasticity problems (Cui et al., 2014; Devincre et al., 2008; Liu et al., 2009; Mishra and Alankar, 2019; Stricker and Weygand, 2015), they remain computationally prohibitive from the viewpoint of bulk deformation studies due to the expense of calculating long-range interactions. Such computational difficulties are perceived to be surmountable with continuum dislocation dynamics, since dislocation-dislocation interactions are computed indirectly via an eigenstrain approach (Xia and El-Azab, 2015; Yefimov et al., 2004).

A typical continuum dislocation dynamics formalism consists of two coupled sets of governing equations (Xia and El-Azab, 2015; Yefimov et al., 2004). The first includes the conventional crystal mechanics equations cast in the form of stress equilibrium and deformation kinematics, with the plastic strain to be fixed from the dislocation motion, and the second comprises the equations governing the transport of and the reactions between dislocations. These two sub-problems are two-way coupled in that the plastic strain is fixed from the dislocation fluxes via Orowan's law, while dislocation motion is dictated by the local internal stress via a dislocation mobility law. A very important aspect of the continuum dislocation dynamics model is the representation of dislocations by multiple density fields distinguished based upon their Burgers vectors and slip planes. Such a representation enables the accurate calculation of both dislocation motion and reactions (El-Azab, 2000; Groma et al., 2003; Hochrainer et al., 2014; Monavari et al., 2016; Monavari and Zaiser, 2018; Sudmanns et al., 2019; Xia and El-Azab, 2015).

The aforementioned representation of dislocations in continuum dislocation dynamics is an extension of the earlier work of Kröner (Kröner, 1958), and Nye (Nye, 1953), who established a theoretical framework that can inform a generic relationship between the dislocation microstructure, the plastic distortion, and the associated internal stress fields in plastically deformed crystals. In such a framework, dislocation density is measured by a dislocation density tensor, $\boldsymbol{\alpha}$, which is defined by the curl of the plastic distortion, $\boldsymbol{\alpha} = -\nabla \times \boldsymbol{\beta}^p$, with the plastic distortion, $\boldsymbol{\beta}^p$, and its elastic counterpart, $\boldsymbol{\beta}^e$, serving as the fundamental components of the deformation of crystals, in the sense that their sum forms the gradient of a compatible displacement field, $\mathbf{u}$. The time evolution of the dislocation density tensor is found from $\dot{\boldsymbol{\alpha}} = \nabla \times (\mathbf{v} \times \boldsymbol{\alpha})$, which was formulated by Mura (Mura, 1963) and Kosevich (Kosevich, 1965). This formula, however, can only be applied to families of dislocations of the same Burgers vector on a relatively small scale since the dislocation velocity $\mathbf{v}$ is only meaningful at that level. In recent years, several attempts have been made to refine an average, statistical description of the evolution of dislocation microstructures. Pioneering work was done for straight, parallel dislocations in two dimensions (2D) by Groma (Groma, 1997), Zaiser *et al* (Zaiser et al., 2001), and Groma *et al* (Groma et al., 2003). The joint evolution of statistically stored and geometric dislocation densities can be captured by these models; see also the recent work of Kooiman *et al* (Kooiman et al., 2014) and Finel *et al* (Cottura et al., 2016). It is quite challenging to extend such 2D approaches to 3D systems where dislocations are interconnected, curved lines moving perpendicular to their line direction. Different approaches have been established to represent 3D dislocation microstructures, for example, by solving for the evolution of the scalar density $\rho(\mathbf{x},t)$ and orientation function $\xi(\mathbf{x},t)$ (Sedláek et al., 2007) or the screw dislocation density, $\rho_{\text{screw}}(\mathbf{x},t)$, and the edge dislocation density, $\rho_{\text{edge}}(\mathbf{x},t)$ (Arsenlis et al., 2004; Grilli et al., 2018; Leung et al., 2015). Another approach proposed by Hochrainer *et al* (Hochrainer et al., 2007) is based on picturing the 3D curved dislocation lines in a higher dimensional space containing line orientation as an additional dimension, so that the dislocation density can carry additional information about the line direction and curvature. Simplified variants of this theory have been formulated, which consider only low-order moments of the dislocation orientation distribution (Sandfeld et al., 2011; Sandfeld and Zaiser, 2015). A further development of this theory is achieved through a hierarchy of evolution equations of the so-called alignment tensors, and the tensors contain information on the directional distribution of dislocation density and dislocation curvature (Hochrainer, 2015; Monavari and Zaiser, 2018). Generally speaking, the implementation of CDD is challenging since the numerical solution of the transport equations always suffers some level of dispersion and diffusion; see for example (Sandfeld et al., 2015; Schulz et al., 2019).

In this paper, a different description of dislocation evolution that was recently formulated by Xia and El-Azab (Xia and El-Azab, 2015) is adopted. In this description, dislocations on various slip systems are represented by vector fields $\boldsymbol{\rho}^{(\alpha)}$, with the superscript being the slip system index. The dislocation tensor $\boldsymbol{\alpha}$ is decomposed according to the contributions of the dislocation vectors $\boldsymbol{\rho}^{(\alpha)}$ from various slip systems weighted by their corresponding crystallographic Burgers vectors, $\mathbf{b}^{(\alpha)}$. Here the vector fields $\boldsymbol{\rho}^{(\alpha)}$ locally represent dislocation bundles with the same line direction. In this case, the direction of the dislocation velocity vector $\mathbf{v}^{(\alpha)}$ for each slip system is uniquely determined as the direction perpendicular to the dislocation line direction.

To complete a full description of crystal plasticity, the stress field applied to the dislocations is needed to fix the dislocation velocity. Different approaches have been adopted in the literature to calculate the long-range interactions arising from dislocation networks. One approach is to develop an analytical expression for the elastic stress between two infinitesimally short dislocation segments based on fundamental dislocation theories (Balluffi, 2016; Hirth and Lothe, 1982). Another approach is to calculate the plastic distortion $\boldsymbol{\beta}^p$ (or the plastic strain $\boldsymbol{\varepsilon}^p$) and use it as an eigenstrain with which to solve the stress equilibrium equation (Evers et al., 2004; Lin et al., 2016, 2015; Xia and El-Azab, 2015; Yefimov et al., 2004). The latter approach is more commonly used in crystal plasticity, because the stress imposed by boundary conditions can also be obtained equivalently by solving the stress equilibrium equation. The evolution of plastic distortion itself is evaluated using Orowan's law, $\dot{\boldsymbol{\beta}}^p = \sum_\alpha -\mathbf{v}^{(\alpha)} \times \boldsymbol{\alpha}^{(\alpha)}$. When the dislocation density field is adequately smooth in space, updating plastic distortion by direct use of Orowan's law provides accurate results. This is the case for most crystal plasticity problems of interest at the mesoscale. However, in this paper, we show that, when dislocation density fluctuates rapidly in space, as in models that aim to capture dislocation patterns, numerical errors in $\boldsymbol{\beta}^p$ can accumulate, resulting in a significant discrepancy between $-\nabla \times \boldsymbol{\beta}^p$ and $\boldsymbol{\alpha}$. We further show that such errors tend to propagate into the predicted stress-strain behavior and dislocation patterns.

In the current work, we use the field dislocation mechanics approach (Acharya, 2004; Acharya and Roy, 2006) in conjunction with our vector density-based dislocation dynamics formulation. The main idea here is to express $\boldsymbol{\beta}^p$ as the sum of two components: $\boldsymbol{\beta}^p = \nabla \mathbf{z} - \boldsymbol{\chi}$, with $\nabla \mathbf{z}$ and $\boldsymbol{\chi}$ being compatible and incompatible components of $\boldsymbol{\beta}^p$, respectively. In this way, the incompatible part $\boldsymbol{\chi}$ is solved for directly from the current dislocation configuration represented by $\boldsymbol{\alpha}$. It is thus guaranteed that the incompatible part of the plastic distortion $\boldsymbol{\beta}^p$ tensor is consistent with the dislocation density tensor $\boldsymbol{\alpha}$, as further explained in the next section.

In Section 2, we introduce the basic formulation of continuum dislocation dynamics in terms of the dislocation vector density and its evolution equations. In Section 3, the algorithms for updating plastic distortion are presented, including the direct time integration algorithm and the field dislocation mechanics formulation. In Section 4, we present the numerical implementation scheme. In Section 5, we present the results produced by the two plastic distortion/stress update algorithms for the case of an austenitic steel crystal under uniaxial loading and multiple slip conditions. We provide brief concluding remarks in the final Section 6.

## 2 Continuum dislocation dynamics model

The concept of distortion is fundamental to the mechanics of deformed crystals (Kröner, 1958). In the case of small deformation, the distortion of a crystal, $\boldsymbol{\beta}$, corresponds to the displacement gradient, $\nabla \mathbf{u}$. This distortion has two components, elastic, $\boldsymbol{\beta}^e$, and plastic, $\boldsymbol{\beta}^p$. We may thus write:

$$\nabla \mathbf{u} = \boldsymbol{\beta} = \boldsymbol{\beta}^e + \boldsymbol{\beta}^p. \tag{1}$$

Due to the presence of dislocations, $\boldsymbol{\beta}^e$ and $\boldsymbol{\beta}^p$ are incompatible fields and thus do not correspond to displacement fields individually. The continuum dislocation dynamics approach is based on the premise that the dislocation system in a deforming crystal can be modeled as a continuous field. To solidify the concepts being discussed, we first consider the case of a single active slip system with all dislocations having the same Burgers vector. With dislocations characterized by two vectors, the line direction and Burgers vector, a dislocation field is often described in the continuum theory by a second order tensor field, $\boldsymbol{\alpha}$, defined in terms of the plastic distortion by (Kröner, 1958):

$$\boldsymbol{\alpha} = -\nabla \times \boldsymbol{\beta}^p. \tag{2}$$

On the other hand, the evolution of the plastic distortion is caused by the motion of dislocations at rate $\dot{\boldsymbol{\beta}}^p$ obtained by Orowan's law,

$$\dot{\boldsymbol{\beta}}^p = -\mathbf{v} \times \boldsymbol{\alpha}, \tag{3}$$

where $\mathbf{v}$ is the velocity field of dislocations. We remark here that the above equation is valid for a single slip situation and at sufficiently small spatial resolution permitting the use of Orowan's law (3). Taking the time derivative of equation (2) and combining it with equation (3), the equation for the time rate of change of the dislocation density tensor $\boldsymbol{\alpha}$ is formed (Mura, 1963),

$$\dot{\boldsymbol{\alpha}} = \nabla \times (\mathbf{v} \times \boldsymbol{\alpha}). \tag{4}$$

The above-stated requirement of sufficiently fine spatial resolution means that, on each slip system the dislocations have the same line direction at a given (continuum) point in space. The dislocation density field then can be represented by a vector field, $\boldsymbol{\rho}$, which carries both the number density and line orientation at every point in space. The dislocation density tensor field $\boldsymbol{\alpha}$ is related to the vector density $\boldsymbol{\rho}$ according to equation (5) (Gurtin et al., 2007)

$$\boldsymbol{\alpha} = \boldsymbol{\rho} \otimes \mathbf{b}, \tag{5}$$

where $\mathbf{b}$ is the Burgers vector of dislocations. In order to obtain the evolution equation of the vector density, $\boldsymbol{\rho}$, equation (5) is substituted into equation (4), and the Burgers vector is dropped, yielding

$$\dot{\boldsymbol{\rho}} = \nabla \times (\mathbf{v} \times \boldsymbol{\rho}), \tag{6}$$

as previously shown by Xia and El-Azab (Xia and El-Azab, 2015). Equation (6) is considered to be the main kinetic equation of dislocation dynamics in our model. It is itself the evolution equation of dislocations on each slip system. It describes the motion and bowing of a dislocation line, and hence the line length increases in a natural way. Also, it can easily be modified to incorporate cross slip by allowing the direction of the velocity vector to change from one slip plane to another. As pointed out in the introduction, that transport equation (6) is based on the line bundle idealization of dislocations (Xia and El-Azab, 2015). This idealization assumes that at a continuum point on a given slip systems, the density has only a single line direction. To satisfy this assumption, the mesh size should be comparable to the annihilation distance of dislocations of opposite sign, and as such the geometric cancellation of dislocations of opposite sign coincides with their physical annihilation.

The vector dislocation density is a combination of the scalar density $\rho$ and the unit tangent to the dislocation line $\boldsymbol{\xi}$,

$$\boldsymbol{\rho} = \rho \boldsymbol{\xi}. \tag{7}$$

The dislocation velocity is also represented by a vector field, written in the form

$$\mathbf{v} = v\boldsymbol{\eta}. \tag{8}$$

As the dislocation line moves perpendicular to its line direction, the dislocation velocity direction $\boldsymbol{\eta}$ can be determined by the slip plane normal $\mathbf{m}$ and the line direction $\boldsymbol{\xi}$ as follows:

$$\boldsymbol{\eta} = \mathbf{m} \times \boldsymbol{\xi}. \tag{9}$$

The scalar velocity $v$ is determined by the resolved shear stress applied on dislocations. According to (Zaiser et al., 2001) and (Yefimov et al., 2004), the dislocation glide velocity is assumed to change linearly with the local resolved shear stress,

$$v = \frac{b}{B}\left\langle \tau - \left(\tau_0 + \tau_p\right)\mathrm{sgn}(\tau) \right\rangle \tag{10}$$

where $b$ is the magnitude of Burgers vector and $B$ is a drag coefficient. The function "sgn" returns the signature of its argument, $\langle \cdot \rangle$ denotes a Macaulay bracket, which returns its argument if it is positive and zero otherwise, $\tau_0$ is the stress representing lattice friction, and $\tau$ is the resolved shear stress corresponding to the Peach-Koehler glide force with magnitude $b\tau$ (Peach and Koehler, 1950). This expression for $v$ includes the long-range interactions between dislocations, together with the stress arising from boundary conditions. Since dislocation reactions, e.g. junctions, can lock or impede the motion of dislocations, a short-range interaction stress is included via $\tau_p$. The details are discussed later.

The dislocation transport equation (6) can be solved when the dislocation velocity field $\mathbf{v}$ is prescribed. As has just been mentioned, the velocity field comes from the resolved shear stress or the Peach-Koehler force according to equation (10), both of which require fixing the stress field, $\boldsymbol{\sigma}$, which is obtained by solving the stress equilibrium equation. The latter equation is part of the crystal mechanics problem, cast below as an eigenstrain problem:

$$\begin{cases} \nabla \cdot \boldsymbol{\sigma} = \mathbf{0} & \text{in } \Omega \\ \boldsymbol{\sigma} = \mathbf{c} : (\nabla \mathbf{u} - \boldsymbol{\beta}^{\text{p}})_{\text{sym}} & \text{in } \Omega \\ \mathbf{u} = \bar{\mathbf{u}} & \text{on } \partial \Omega_u \\ \mathbf{n} \cdot \boldsymbol{\sigma} = \bar{\mathbf{t}} & \text{on } \partial \Omega_\sigma \end{cases}, \tag{11}$$

in which $\mathbf{c}$ is the elastic tensor. In the above, the plastic distortion (or its symmetric part, the plastic strain) is considered as the eigenstrain. The numerical update of eigenstrain (eigendistortion) represents the main motivation of this work and it is explained in Section 3. In equation (11), $\Omega$ is the crystal domain of the problem and its boundary is $\partial \Omega$, with the latter partitioned into parts $\partial \Omega_u$ over which the displacement boundary condition $\bar{\mathbf{u}}$ is applied, and $\partial \Omega_\sigma$ over which a traction boundary condition $\bar{\mathbf{t}}$ is applied, where $\mathbf{n}$ is the unit outward normal to the boundary. By solving equation (11), the displacement and stress fields can be obtained from both imposed systems of boundary tractions and internal dislocations (eigenstrain).

The Peach-Koehler force, $\mathbf{f}_{\text{PK}}$, acting on the dislocations in a stress field, $\boldsymbol{\sigma}$, is given by (Hirth and Lothe, 1982)

$$\mathbf{f}_{\text{PK}} = (\mathbf{b} \cdot \boldsymbol{\sigma}) \times \boldsymbol{\xi}, \tag{12}$$

where $\mathbf{f}_{\text{PK}}$ is the Peach-Koehler force per unit length of dislocation, $\mathbf{b}$ is the Burgers vector, and $\boldsymbol{\xi}$ is the local tangent of the dislocation bundle. Equation (12) can be rewritten to obtain the glide component and the climb component. Based on the definition of the velocity direction $\boldsymbol{\eta}$ (equation (9)), we have $\boldsymbol{\xi} = \boldsymbol{\eta} \times \mathbf{m}$. So

$$\mathbf{f}_{\text{PK}} = (\mathbf{b} \cdot \boldsymbol{\sigma}) \times (\boldsymbol{\eta} \times \mathbf{m}) = (\mathbf{b} \cdot \boldsymbol{\sigma} \cdot \mathbf{m}) \boldsymbol{\eta} - (\mathbf{b} \cdot \boldsymbol{\sigma} \cdot \boldsymbol{\eta}) \mathbf{m}. \tag{13}$$

Equation (13) shows the glide component and climb component of the Peach-Koehler force clearly,

$$\mathbf{f}_g = b(\mathbf{s} \cdot \boldsymbol{\sigma} \cdot \mathbf{m}) \boldsymbol{\eta} \quad \text{and} \quad \mathbf{f}_c = -b(\mathbf{s} \cdot \boldsymbol{\sigma} \cdot \boldsymbol{\eta}) \mathbf{m}. \tag{14}$$

In the above, $\mathbf{s}$ is the unit slip direction $\mathbf{s} = \mathbf{b}/|\mathbf{b}|$. Only the glide component of the motion is considered in the dislocation transport, which corresponds to a resolved shear stress $\tau$ of the form

$$\tau = \mathbf{s} \cdot \boldsymbol{\sigma} \cdot \mathbf{m}, \tag{15}$$

to be used in equation (10).

A generalization to a multi-slip situation requires equations (6), (10), (13) and (15) for the dislocation transport, dislocation velocity, Peach-Koehler force and resolved shear stress, respectively, to be specialized to individual slip systems. This in turn requires the dislocation density, $\boldsymbol{\rho}$, velocity, $\mathbf{v}$, slip plane normal, $\mathbf{m}$, and slip direction, $\mathbf{s}$, Burgers vector, $\mathbf{b}$, resolved shear stress, $\tau$, Peach-Koehler force, $\mathbf{f}_{PK}$, and its glide and climb components, and the short range stress, $\tau_p$, to be so specified. We will do so by adding superscripts, $\alpha$, $\beta$, etc.

As dislocations evolve, dislocation reactions and cross slip take place. In the current formulation, some reactions are explicitly considered such as the glissile junction formation and collinear annihilation. Others such as the immobile Lomer and Hirth junctions, which contribute part of the hardening behavior, are modeled via a short-range, Taylor type term, $\tau_p$, in the mobility law, Eq. (10). The form of this term is taken after (Devincre et al., 2006; Franciosi et al., 1980; Kubin et al., 2008):

$$\tau_p^\alpha = \mu b \sqrt{a^{\alpha\beta} \rho^\beta}, \qquad (16)$$

with $\mu$ being the shear modulus and $a^{\alpha\beta}$ an interaction coefficient representing the average strength of the interaction between slip systems $\alpha$ and $\beta$.

Dislocation cross slip, collinear annihilation, and glissile junction reactions, influence the balance of dislocations on various slip systems and thus lead to additional terms in the dislocation transport equation (Eq. (6)). The latter thus is modified by adding an extra term, $\dot{\boldsymbol{\rho}}_{cp}^\alpha$, representing the contributions of these processes. That is,

$$\dot{\boldsymbol{\rho}}^\alpha = \nabla \times \left( \mathbf{v}^\alpha \times \boldsymbol{\rho}^\alpha \right) + \dot{\boldsymbol{\rho}}_{cp}^\alpha. \qquad (17)$$

Cross slip and reactions change the evolution of a dislocation network and lead to dislocation multiplication. The cross slip and collinear annihilation terms with the same Burgers vector will be coupled. Glissile junction reactions couple three slip systems, say, $\alpha$, $\beta$, $\gamma$, where Burgers vectors satisfy $\mathbf{b}^\alpha + \mathbf{b}^\beta = \mathbf{b}^\gamma$. A detailed formulation of these coupling terms can be found in an earlier work by the current authors (Lin and El-Azab, 2020).

## 3 Plastic distortion and stress update algorithms

We now turn to the update algorithm for the plastic distortion tensor, $\boldsymbol{\beta}^p$, which is required to calculate the long-range stress on the dislocation system subject to the prescribed boundary conditions. As discussed earlier, $\boldsymbol{\beta}^\mathrm{p}$ serves as an eigenstrain in the crystal mechanics problem described in equation (11). The dislocation dynamics problem yields the rate of the plastic distortion in terms of Orowan's law, equation (3), which upon combining with equations (7) through (9) and considering a multi-slip situation, gives

$$\dot{\boldsymbol{\beta}}^\mathrm{p} = -\sum_\alpha v^\alpha \boldsymbol{\eta}^\alpha \times \left(\rho^\alpha \boldsymbol{\xi}^\alpha \otimes \mathbf{b}^\alpha\right) = \sum_\alpha v^\alpha \rho^\alpha b^\alpha \mathbf{m}^\alpha \otimes \mathbf{s}^\alpha = \sum_\alpha \dot{\gamma}^\alpha \mathbf{m}^\alpha \otimes \mathbf{s}^\alpha , \qquad (18)$$

with $\dot{\gamma}^\alpha = v^\alpha \rho^\alpha b^\alpha$ being the slip rate or the time derivative of the local shear strain $\gamma^\alpha$ on slip system $\alpha$. We remark here that, in crystal plasticity models of small strain, the plastic distortion $\boldsymbol{\beta}^\mathrm{p}$ is expressed in terms of crystal slips $\gamma^\alpha$ as follows:

$$\boldsymbol{\beta}^\mathrm{p} = \sum_\alpha \gamma^\alpha \mathbf{m}^\alpha \otimes \mathbf{s}^\alpha . \qquad (19)$$

*Algorithm 1: Update of $\boldsymbol{\beta}^\mathrm{p}$ by direct time integration of $\dot{\boldsymbol{\beta}}^\mathrm{p}$:* From equations (18) and (19), it can be seen that, to update the plastic distortion tensor $\boldsymbol{\beta}^\mathrm{p}$, it is only necessary to calculate the crystal slip magnitudes $\gamma^\alpha$ for all slip systems by integrating the corresponding slip rates. That is,

$$\gamma^\alpha(\mathbf{x}) = \int_0^t \dot{\gamma}^\alpha(\mathbf{x}, t') dt' . \qquad (20)$$

The plastic distortion can then be constructed by inserting all $\gamma^\alpha$ values from equation (20) into equation (19). As mentioned in the introduction, this algorithm results in an inconsistency between the dislocation density tensor, which is the measure of the elastic strain incompatibility, and the curl of the updated plastic distortion. This inconsistency arises due to accumulation of the numerical errors in the staggered solution schemes of the coupled dislocation transport/crystal mechanics equations. Such errors result in a stress field that is inconsistent with the dislocation tensor itself, the latter being the origin of stress. The same inconsistency also appears in differences between the slip system dislocation density and the curls of the 'oriented' plastic slip (shear strain) on individual slip systems. The latter is the plastic shear strain taken together with the unit normal to the slip plane for a given slip system.

*Algorithm 2: Update of $\boldsymbol{\beta}^\mathrm{p}$ using the field dislocation mechanics approach:* In order to obtain the stress field accurately, we adopt the field dislocation mechanics formulation (Acharya and Roy, 2006; Brenner et al., 2014; Roy and Acharya, 2006, 2005) to update the plastic distortion tensor.

In field dislocation mechanics, the plastic distortion $\boldsymbol{\beta}^p$ is divided into a compatible part $\nabla \mathbf{z}$ and an incompatible part $\boldsymbol{\chi}$,

$$\boldsymbol{\beta}^p = \nabla \mathbf{z} - \boldsymbol{\chi}. \tag{21}$$

Being the gradient of a vector field $\mathbf{z}$, the compatible part is curl-free. The incompatible part, on the other hand is, divergence-free. The field problem governing the incompatible part, $\boldsymbol{\chi}$, can be derived from equations (2) and (21),

$$\begin{cases} \nabla \times \boldsymbol{\chi} = \boldsymbol{\alpha} & \text{in } \Omega \\ \nabla \cdot \boldsymbol{\chi} = 0 & \text{in } \Omega \\ \mathbf{n} \cdot \boldsymbol{\chi} = 0 & \text{on } \partial \Omega \end{cases} \tag{22}$$

The vector field $\mathbf{z}$ associated with the compatible part of $\boldsymbol{\beta}^p$ satisfies a boundary value problem derived from Eqs. (3) and (21),

$$\begin{cases} \nabla \cdot \nabla \dot{\mathbf{z}} = \nabla \cdot (-\mathbf{v} \times \boldsymbol{\alpha}) & \text{in } \Omega \\ \mathbf{n} \cdot \nabla \dot{\mathbf{z}} = \mathbf{n} \cdot (-\mathbf{v} \times \boldsymbol{\alpha}) & \text{on } \partial \Omega \\ \dot{\mathbf{z}} = \dot{\mathbf{z}}_o \, (\text{arbitrary value}) & \text{at one point in } \Omega \end{cases} \tag{23}$$

The boundary conditions in equations (22) and (23) are generally applicable and have been discussed in detail by Acharya (Acharya and Roy, 2006). They are used to obtain a unique decomposition of the plastic distortion tensor.

Because we concern ourselves with the total plastic distortion, we first assemble the dislocation density tensor, $\boldsymbol{\alpha} = \sum_{\alpha} \boldsymbol{\rho}^{\alpha} \otimes \mathbf{b}^{\alpha}$. We then use it in equation (22) and we substitute the Orowan's term, $\sum_{\alpha} \mathbf{v}^{\alpha} \times \boldsymbol{\alpha}^{\alpha}$, into Eq (23). Upon solving the last two boundary value problems for the compatible and incompatible parts of the plastic distortion tensor, $\boldsymbol{\beta}^p$, the latter can be assembled as per equation (21). We remark here that, although the compatible part still needs integration over time, the incompatible part will be calculated directly from the current dislocation density tensor regardless of the history of dislocation evolution. We will show the importance of calculating the incompatible part independently of the evolution history in a test example later.

It should be pointed out that only the decomposition of plastic distortion of Acharya and Roy's field dislocation mechanics formalism (Acharya and Roy, 2006) into compatible and incompatible parts is adopted here. Our dislocation transport equations (Eq. (17)) are different. While Acharya and Roy use a second-order dislocation density tensor to represent the dislocation field, a set of slip-system level dislocation vector densities are used in our case. This enables us to compute the dislocation velocity from a Peach-Koehler force via a mobility law, and it enables us to explicitly incorporate cross slip and reactions along with their line direction dependence.

## 4 Numerical Solution Scheme

As presented above, the overall problem of continuum dislocation dynamics can be divided into two parts, solving the stress equilibrium equation (11) for the displacement field $\mathbf{u}$ given the plastic distortion distribution, together with the dislocation kinetic equations (6) or (17) for the set of $\rho^\alpha$. With Algorithm 2 for the plastic distortion update, two additional sets of boundary value problems, summarized in equations (22) and (23), are to be solved for $\mathbf{z}$ and $\boldsymbol{\chi}$. The latter can be considered as subsidiary problems to be solved for the proper update of the plastic distortion and, in turn, the stress tensor. We employ a staggered scheme to solve the coupled crystal mechanics/dislocation kinetics problem. We first assign initial densities to all slip systems within the simulation domain with an initially known plastic distortion field $\boldsymbol{\beta}^\mathrm{p}$. We next solve equation (22) and (23) to obtain the corresponding plastic distortion $\boldsymbol{\beta}^\mathrm{p}$. Then the stress field can be obtained by using the stress equilibrium equation (11). Once the stress field is found, the resolved shear stress equation (15) and the dislocation velocity equation (10) can be used to solve the kinetic equation for dislocation density evolution, per equation (17). The new dislocation configuration is used for the next time step until the average strain reaches a desired value. The overall scheme is schematically shown in Figure 1.

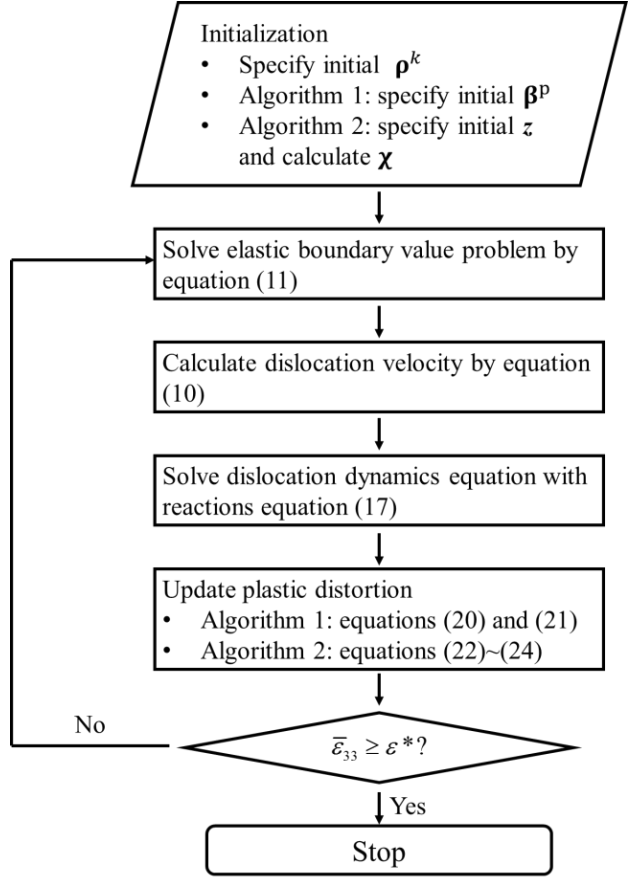

Figure 1. The flowchart for the solution scheme. $\bar{\varepsilon}_{33}$ is the applied strain during loading and $\varepsilon^*$ is the final strain.

We use a standard Galerkin finite element method to solve the stress equilibrium equation and the boundary value problem for the compatible part of plastic distortion, while, for ease of enforcing the divergence-free constraints, the least squares finite element method is used for the kinetic equation for dislocation density and the incompatible part of plastic distortion. Details of the finite element formulations of these equations are in Appendix.

## 5 Results and discussions

To illustrate the improvement in the manner of updating plastic distortion by using field dislocation mechanics instead of direct time integration of a rate expression, we show a test example in which two dislocation loops expand with a prescribed velocity and then annihilate with each other. Then a bulk simulation for an FCC crystal (austenitic steel) with [001] loading, performed by the continuum dislocation dynamics with field dislocation mechanics, is shown to illustrate the microstructure evolution during loading.

**5.1 Expansion and annihilation of two dislocation loops**

The two alternative algorithms for updating plastic distortion are compared in this example. Initially, two dislocation bundles in the form of loops were placed close together in a simulation domain of size 5µm×10µm×5.303µm, as shown in Figure 2. Dislocation line bundles are represented by a vector field initialized by a Gaussian distribution function over their cross section. The values in Figure 2 show the norm of dislocation density tensor $\boldsymbol{\alpha}$, which is found from the dislocation density vector $\boldsymbol{\rho}^{(k)}$ and Burger vector $\mathbf{b}^{(k)}$ via Eq. (5). The $x$, $y$ and $z$ axes are taken to be along [110], [$\bar{1}$10] and [001], respectively, so that the dislocation loops are on a (111) slip plane. The Burgers vector is oriented along the $y$ axis. A hybrid mesh with tetrahedral and pyramidal elements is used to replicate the FCC lattice structure, as described in (Xia and El-Azab, 2015). The mesh size here is $l_{mesh} = 0.125$ µm. A prescribed dislocation velocity $v = 0.03$ µm/ns is used to cause the two loops to expand and the time step $\Delta t = 1.875$ ns. Periodic boundary conditions are used in this test example.

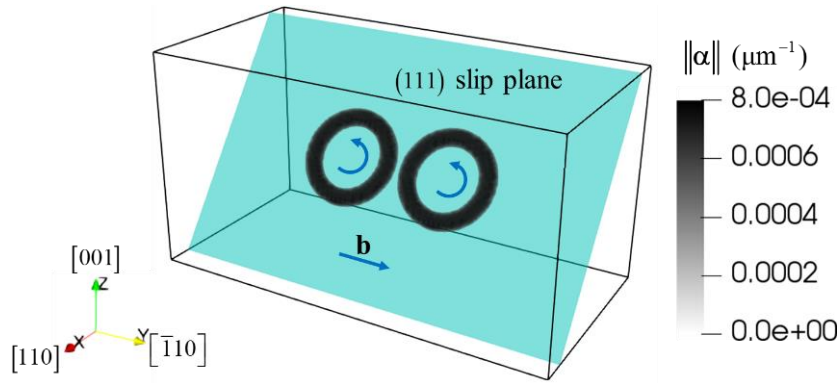

Figure 2 Initial dislocation structure. Two dislocation loops are placed on a (111) slip plane.

The evolution of the dislocation density, calculated by solving the transport equation (17), is depicted in Figure 3 in a view perpendicular to the slip plane. Aside from some diffusion, the evolution of the two dislocation loops looks reasonably accurate. At first, the two loops expand and then they interact with one another and begin to annihilate in places where opposing line directions intersect. Finally, the two dislocation loops merge into one single dislocation loop that continues to expand. This simple result shows that the dislocation transport-reaction equation is solved accurately. In particular, by using the least squares finite element method and applying a divergence-free constraint on the dislocation density vector, the annihilation process is captured without the numerical dispersion that emerges whenever first-order hyperbolic partial differential equations are not carefully solved (Varadhan et al., 2006).

It is to be noted here that, in this test problem, the dislocation transport equation (Eq. (6)) is solved using the same prescribed dislocation velocity for both algorithms. As such, the evolution of the dislocation field is the same, which facilitates the comparison of the plastic distortion fields resulting from the two update algorithms.

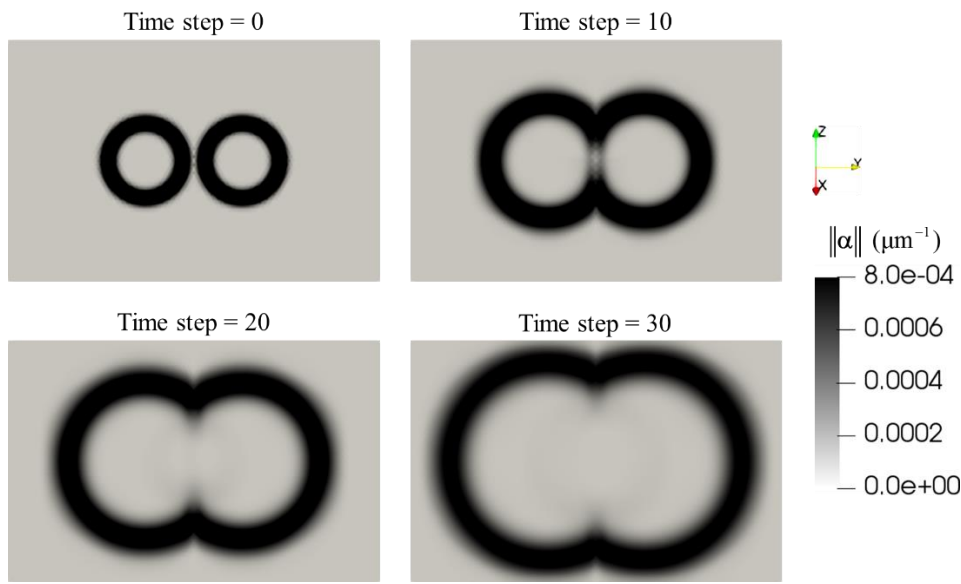

Figure 3 Dislocation density evolution on the slip plane. The two loops expand and annihilate with each other to form a larger loop.

As the dislocation density evolves, the associated plastic distortion will also change. As discussed in Section 3, two algorithms were used for comparison to update the plastic distortion. The plastic distortion calculated by direct time integration of the rate expression fixed by Orowan's law is denoted as $\beta_I^p$ and the one found by our field dislocation mechanics approach is denoted as $\beta_{II}^p$. The evolutions of these two quantities are shown in Figure 4 and Figure 5. (A Burgers vector of 0.25 nm was used). The slipped area inside dislocation loops is shown in red, while the blue area shows unslipped area outside the dislocation loops. Figure 4 and Figure 5 can be compared with Figure 3 via the slipped area. The slipped area in Figure 3 is the area inside the dislocation loops. It can be seen in Figure 3 that the slipped area increases as the dislocation loops expand. After parts of the loops encounter each other and annihilate, the slipped area becomes connected. The slip area in Figure 3 is proportional to the plastic shear strain on the slip system considered in this simulation, and the latter gives the magnitude of the plastic distortion $\boldsymbol{\beta}^\text{p}$. As such, the numerical algorithm that yields a plastic distortion distribution consistent with the slip area in Figure 3 is the more accurate one. Returning to Figure 4 and Figure 5, we notice that there are differences between the spatial distortion distributions of $\boldsymbol{\beta}_\text{I}^\text{p}$ and $\boldsymbol{\beta}_\text{II}^\text{p}$. For example, at time steps 20 and 30 in Figure 4, fluctuations can be seen at the center where dislocations annihilate. It seems that the direct time integration algorithm accumulates error in calculating the plastic distortion in regions where dislocations of opposite direction annihilate. By using field dislocation mechanics (as shown in Figure 5), the evolution of the plastic distortion is similar, except that the fluctuations are highly reduced at the center. As shown later, the fluctuations in the field dislocation mechanics simulation only arise from the time integration of the compatible part, while the incompatible part is consistent with the current dislocation density tensor. This enhanced numerical stability has important implications for stress calculations.

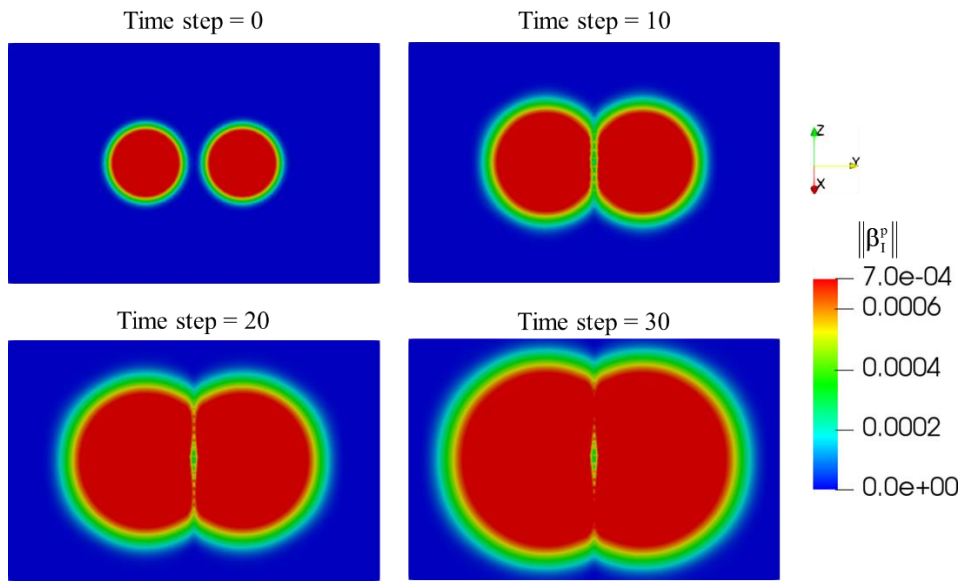

Figure 4 Evolution of the plastic distortion updated using the direct time integration algorithm.

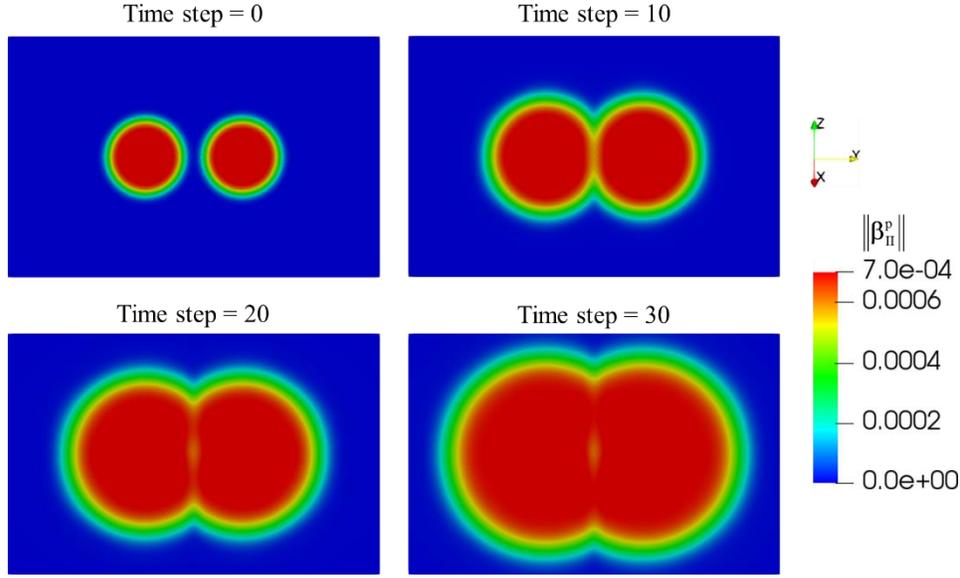

Figure 5 Evolution of the plastic distortion, updated using the field dislocation mechanics algorithm.

To see how accurately the plastic distortion is updated, we take the numerical curl of the plastic distortion field updated by the two algorithms, $\boldsymbol{\alpha}_\mathrm{I} = -\nabla \times \boldsymbol{\beta}_\mathrm{I}^\mathrm{p}$ and $\boldsymbol{\alpha}_\mathrm{II} = -\nabla \times \boldsymbol{\beta}_\mathrm{II}^\mathrm{p}$, where $\boldsymbol{\alpha}_\mathrm{I}$ and $\boldsymbol{\alpha}_\mathrm{II}$ are the dislocation density tensors related to the corresponding plastic distortion field. Their evolutions are shown in Figure 6 and Figure 7. The dislocation density tensor $\boldsymbol{\alpha}$, obtained from solving the transport equation, was already shown in Figure 3. Comparing $\boldsymbol{\alpha}_\mathrm{I}$ and $\boldsymbol{\alpha}_\mathrm{II}$ from Figure 6 and Figure 7, respectively, with the $\boldsymbol{\alpha}$ of Figure 3, we can see the differences in accuracy between the two algorithms. By using direct time integration, the dislocation density tensor $\boldsymbol{\alpha}_\mathrm{I}$ noticeably differs from $\boldsymbol{\alpha}$. Namely, there are dislocations left in the annihilation region (Figure 6), which should not exist (Figure 3). This means that, although the dislocation evolution equation (17) is solved accurately in our model, the curl of the plastic distortion $\boldsymbol{\beta}^\mathrm{p}$ may be inconsistent with the dislocation density tensor if the plastic distortion is updated by direct time integration of the rate expression derived from Orowan's law. Numerical errors can accumulate during the numerical integration operation. Another difference between the dislocation density distributions of Figure 6 and Figure 3 is that there are two small dislocation loops with small density values in Figure 6 at which the initial loops are placed. Although an analytical expression of the initial plastic distortion $\boldsymbol{\beta}_\mathrm{I}^\mathrm{p}$ is consistent with the analytical expression of the initial dislocation field $\boldsymbol{\alpha}$, the numerical curl operation from values at nodes of the mesh may lead to inconsistency in the initial values of the two fields. This inconsistency remains for the subsequent time steps. On the other hand, by using field dislocation mechanics, these problems are eliminated. The dislocation density tensor in Figure 7 is almost the same as in Figure 3. Since the incompatible part of $\beta_{II}^{P}$ is calculated directly from the current dislocation density tensor $\alpha$, numerical error is much smaller.

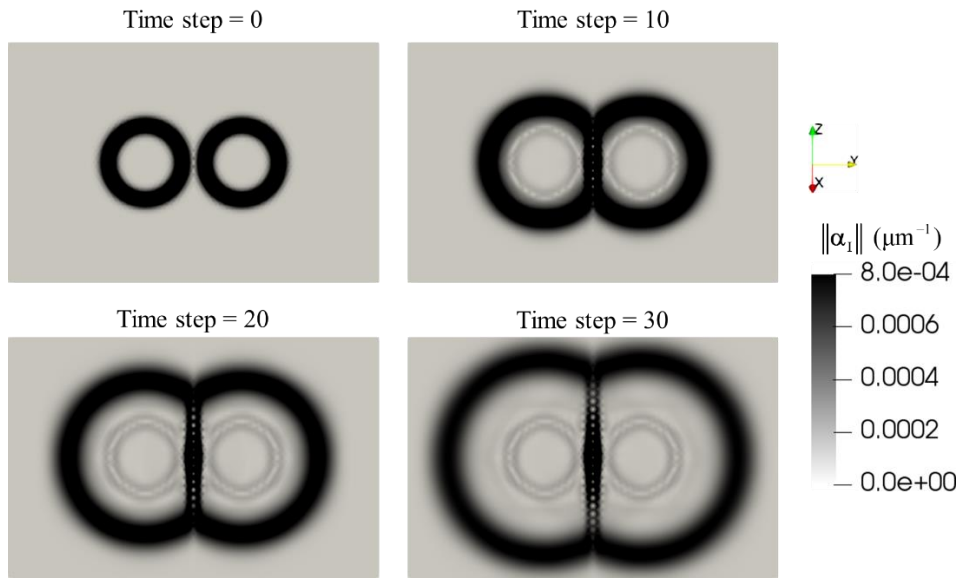

Figure 6 The dislocation density tensor calculated from the plastic distortion updated by direct time integration.

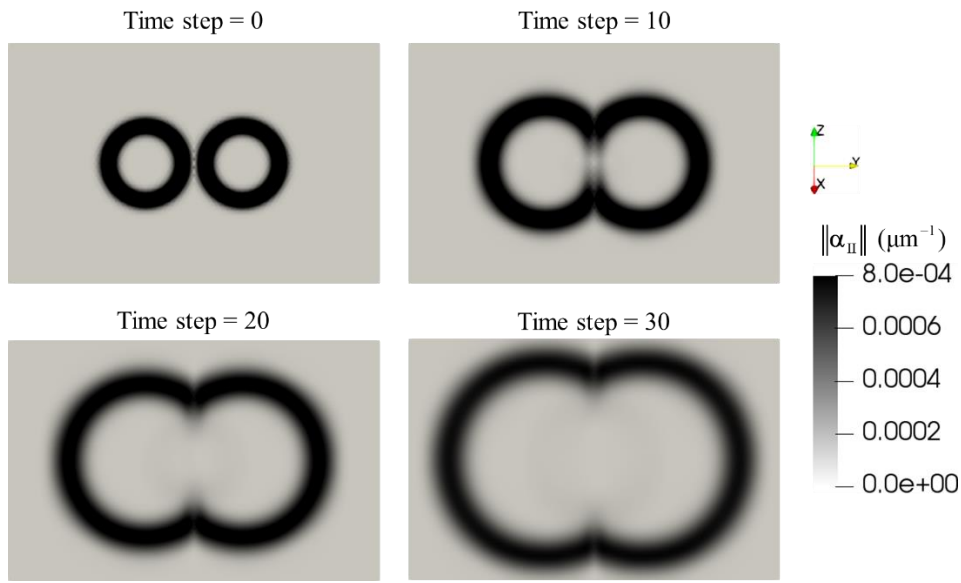

Figure 7 The dislocation density tensor calculated from plastic distortion, as updated by field dislocation mechanics.

In order to compare the accuracy of the update two algorithms quantitatively, it will be required to compare the errors in the plastic distortion obtained by the two algorithms relative to the plastic distortion computed by a third and a more accurate method. Generally speaking, such a third solution is not available. For the test examples presented here, however, that third method would be the analytical solution of the plastic distortion, which can in principle be found from knowledge of the density and the assumed velocity of dislocations. Such a comparison, however, would test together the accuracy of the solution of the transport equations together with the update algorithm, which is not the objective here. An alternate test of accuracy of the update algorithms is the error of the dislocation density computed as the curl of the plastic distortion relative to that which comes from the solution of the transport equation. Let us denote the latter by $\boldsymbol{\alpha}_I$ and $\boldsymbol{\alpha}_{II}$ for the two update algorithms. The relative error of interest, $\alpha_{\text{Error}}$, can be defined in an integral sense over the domain of the solution by

$$\alpha_{\text{Error, I(II)}} = \frac{\int_\Omega \|\boldsymbol{\alpha}_{I(II)} - \boldsymbol{\alpha}\| \, dV}{\int_\Omega \|\boldsymbol{\alpha}\| \, dV}, \tag{24}$$

where $\Omega$ is the volume of the simulation domain and $\|\cdot\|$ is the norm operator for a second order tensor. The evolution of $\alpha_{\text{Error}}$ is shown in Figure 8. The figure shows that the direct time integration algorithm yields a much higher numerical error than that of the field dislocation mechanics approach. The rate of increase of $\alpha_{\text{Error}}$ is also higher in the case of direct time integration. Ideally speaking, the field dislocation mechanics approach should not accumulate errors. However, as the way the test is performed here is that we are comparing the gradients of the plastic distortion itself, which is obtained by numerical differentiation, i.e., by taking the numerical curl $\boldsymbol{\alpha}_I = -\nabla \times \boldsymbol{\beta}_I^p$ and $\boldsymbol{\alpha}_{II} = -\nabla \times \boldsymbol{\beta}_{II}^p$, which introduce errors.

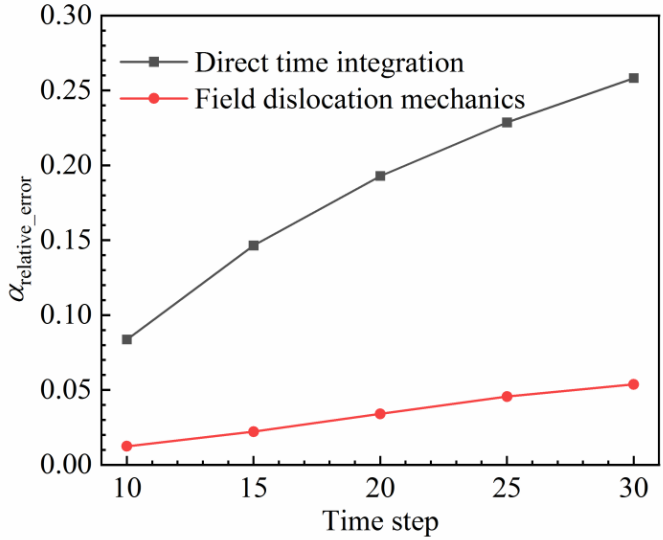

Figure 8 The relative error of dislocation density tensor (defined by Eq. (24)) calculated by the two algorithms.

Although the dislocation velocity is prescribed and not calculated from the stress field in this test example, the stress fields $\boldsymbol{\sigma}_\mathrm{I}$ and $\boldsymbol{\sigma}_\mathrm{II}$, obtained by using $\boldsymbol{\beta}_\mathrm{I}^\mathrm{p}$ and $\boldsymbol{\beta}_\mathrm{II}^\mathrm{p}$ as eigendistortions, can be compared to show how dislocation evolution would be affected if the stress field were used. The stress field is calculated using periodic boundary conditions, and the average stress is zero. The resolved shear stress values, $\tau_\mathrm{I} = \mathbf{m}\cdot\boldsymbol{\sigma}_\mathrm{I}\cdot\mathbf{s}$ and $\tau_\mathrm{II} = \mathbf{m}\cdot\boldsymbol{\sigma}_\mathrm{II}\cdot\mathbf{s}$ at time step 30 are shown in Figure 9. A Young's modulus value of 112.5 GPa, a Poisson ratio of 0.34, and a Burgers vector of 0.25 nm were used in this stress calculation test. Figure 9 clearly shows that an inaccurate plastic distortion will lead to a significant error in the stress field. The spurious dislocations shown in Figure 6 cause the negative resolved shear stress at the center (left in Figure 9). The small dislocation loops in Figure 6 also perturb the stress field. On the other hand, the resolved shear stress calculated from field dislocation mechanics is more representative of the current dislocation state. It only has non-zero values around dislocation bundles, and it changes sign across the bundles. An accurate stress field is very important for showing correct patterns in bulk plasticity simulations at mesoscale, in which there are many dislocations and dislocation annihilation events occurring simultaneously.

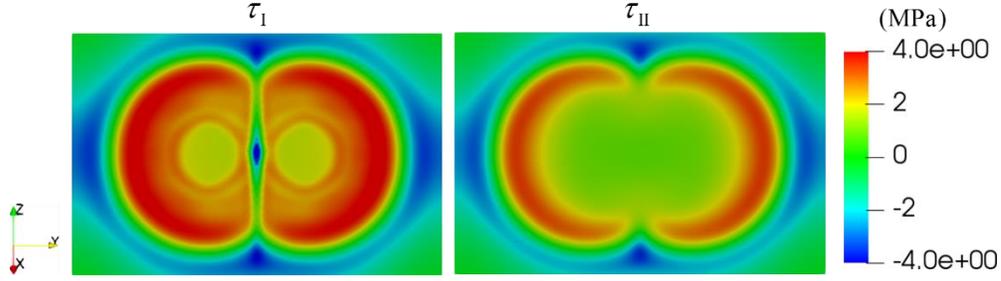

Figure 9 The resolved shear stress, as alternately calculated from direct time integration (left) and field dislocation mechanics (right).

As has been shown above, a direct time integration of the plastic distortion may lead to an inaccurate stress field. Accuracy can be improved by using field dislocation mechanics, in which the incompatible part of the plastic distortion is calculated directly from the current dislocation density tensor. However, the compatible part of the plastic distortion still needs time integration, see equation (23). A question that arises here is the extent to which the accuracy of this time integration affects the stress field. The stress field is not affected at all, if all mechanical boundaries are traction boundaries. Suppose the exact plastic distortion $\boldsymbol{\beta}^\mathrm{p}$ has a compatible part $\nabla \mathbf{z}$ and an incompatible part $\boldsymbol{\chi}$, the compatible part of the plastic distortion $\boldsymbol{\beta}^\mathrm{p}_\mathrm{II}$ calculated from field dislocation mechanics differs from the exact solution $\nabla \mathbf{z}_\mathrm{II} = \nabla(\mathbf{z} + \delta\mathbf{z})$, while the incompatible part is exact $\boldsymbol{\chi}_\mathrm{II} = \boldsymbol{\chi}$. Then the corresponding displacement and stress fields satisfy the following two sets of equations,

$$\begin{cases} \nabla \cdot \boldsymbol{\sigma} = \mathbf{0} & \text{in } \Omega \\ \boldsymbol{\sigma} = \mathbf{C} : (\nabla \mathbf{u} - \nabla \mathbf{z} + \boldsymbol{\chi})_\mathrm{sym} & \text{in } \Omega \\ \mathbf{n} \cdot \boldsymbol{\sigma} = \bar{\mathbf{t}} & \text{on } \partial\Omega_\sigma \end{cases} \quad (25)$$

$$\begin{cases} \nabla \cdot \boldsymbol{\sigma}_\mathrm{II} = \mathbf{0} & \text{in } \Omega \\ \boldsymbol{\sigma}_\mathrm{II} = \mathbf{C} : (\nabla \mathbf{u}_\mathrm{II} - \nabla \mathbf{z} - \nabla \delta\mathbf{z} + \boldsymbol{\chi})_\mathrm{sym} & \text{in } \Omega \\ \mathbf{n} \cdot \boldsymbol{\sigma}_\mathrm{II} = \bar{\mathbf{t}} & \text{on } \partial\Omega_\sigma \end{cases} \quad (26)$$

If $\mathbf{u}$ is the solution of equation (25), it is easy to prove that

$$\mathbf{u}_\mathrm{II} = \mathbf{u} + \delta\mathbf{z} \quad (27)$$

is the solution of equation (26). Substituting it back into equation (26), it is easy to obtain the following result,

$$\boldsymbol{\sigma}_{II} = \boldsymbol{\sigma}. \tag{28}$$

Equations (27) and (28) imply that if the compatible part of the plastic distortion is calculated inaccurately, an error results in the displacement field, while the stress field is not affected. This fact explains why there are plastic distortion fluctuations at the center of Figure 5, but no stress fluctuation are present at the center of Figure 9 (right), which was generated from field dislocation mechanics calculations. All these errors derive from the calculation of the compatible part of the plastic deformation. So as long as an exact solution of the incompatible part of the plastic distortion is obtained, the stress field is exact. This fidelity can be guaranteed by using field dislocation mechanics, but not by using direct time integration. Also, although updating the compatible part of the plastic distortion involves time integration, this procedure incurs smaller errors than integrating the whole plastic distortion tensor over time, as per Figure 4 and Figure 5. It is noted that equations (27) and (28) hold only when all boundaries are traction boundaries. If a displacement boundary condition is applied, then the stress field calculated will be slightly different from the accurate value at that boundary, but it will remain accurate in the domain far from the displacement boundary.

## 5.2 Bulk simulation with multiple slip systems

The purpose of this section is to present a larger test problem for the application of the update algorithms by running a 3D simulation with multiple slip in FCC crystal. In some simulations involving direct time integration, solution instability occurred after about $10^5$ time steps, which is significantly before the target strain range is reached. That instability was manifested in the form of unbounded stress and dislocation density fluctuation in localized regions in the domain of solution. However, by using the field dislocation mechanics, the stress field is updated much more accurately and the solution remains stable well past a million time steps. As such, only the update algorithm based on field dislocation mechanics is used in the test presented in this section.

The simulation domain used is a 5µm×5µm×5.303µm box, as shown in Figure 10(a). The *x*, *y* and *z* axes are selected to be along the [110], [$\bar{1}$10] and [001] crystallographic directions, respectively. The domain is discretized using a hybrid mesh with pyramidal and tetrahedral elements. The mesh size is chosen to be 62.5nm, so that the annihilation distance for an edge dislocation dipole is consistent with the property of the material (25 nm). Initial dislocation structures are set by placing dislocation loops in the domain with periodic boundaries, meaning that if a dislocation exits the domain through one surface, it re-enters the domain from the opposite surface. In this simulation, 10 dislocation loops were placed randomly on each of the 12 slip systems of an FCC crystal, with radii ranging from 2µm to 6 µm, as shown in Figure 10(b). The initial total dislocation density over the domain is scaled to 1.5 µm$^{-2}$. Multiple dislocation interactions were considered. Cross slip was included with a coarse-grained rate found from discrete dislocation dynamics (DDD) simulations (Xia et al., 2016). Collinear annihilation and glissile junctions were considered in order to account for dislocation exchange between different slip systems. Lomer-Cottrell and Hirth locks were incorporated via a Taylor hardening term, with coefficients 0.084 (Lomer-Cottrell) and 0.051 (Hirth) (Madec, 2003).

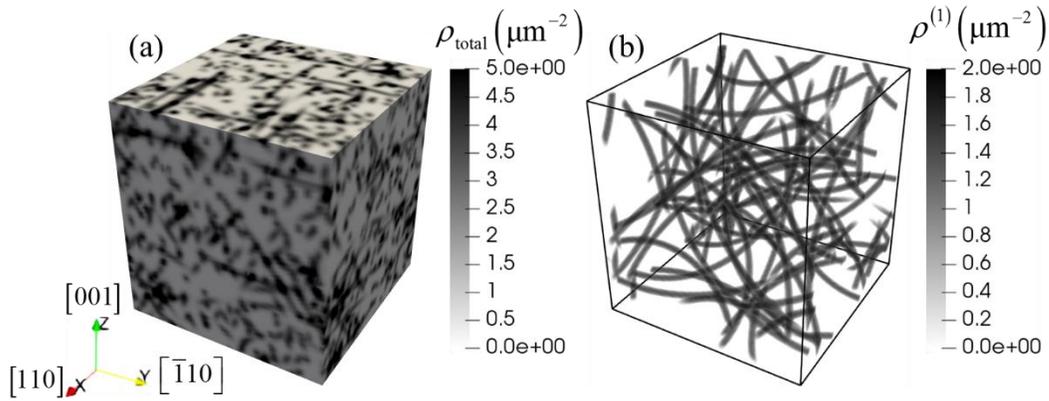

Figure 10 Initial dislocation structure. Each slip system has 10 dislocation loops placed randomly in the domain with radii ranging from 2 µm to 6 µm. (a) Total dislocation density; (b) Dislocation density on the first slip system.

Material parameters were chosen to be those of steel (Déprés et al., 2004), as shown in Table 1. The crystal was loaded along the [001] direction at a strain rate of 20 s$^{-1}$. All boundaries are as periodic boundaries. An adaptive time step scheme was used, which was controlled by a Courant number $c = v_{max} \Delta t / l_{mesh} = 0.45$, with $v_{max}$ being the maximum dislocation velocity in the domain. Typically, the calculated time increment for each step is about several nanoseconds. A relaxation step is performed prior to loading to release the stress field introduced by placing the initial dislocations into the simulation domain. It is performed by letting the initial dislocation microstructure evolve under zero applied stress until the maximum dislocation velocity in the is very close to zero, typically less than 10$^{-2}$ m/s.

Table 1 Material parameters used in the simulation (steel) (Déprés et al., 2004).

| Young's modulus (GPa) | Possion's ratio | Burgers vector (nm) | Drag coefficient (Pa·s) | Lattice friction stress (MPa) |
|---|---|---|---|---|
| 189 | 0.26 | 0.254 | $7.12 \times 10^{-6}$ | 3 |

The crystal is loaded to 1.5% strain. The stress-strain curve is shown in Figure 11. It starts in the elastic regime, which extends to 0.1% strain. A clear yield point of about 13 MPa is shown. After dislocations start to move, the crystal deformation progresses into the plastic regime. The hardening rate is roughly a constant after 1% strain. [001] loading is highly symmetric, with eight active slip systems. The reactions between these slip systems are the origin of the strain hardening. At 1.5% strain, the applied stress reaches about 17 MPa. The evolution of the dislocation density is shown in Figure 12. The total dislocation density increases with strain (Figure 12(a)). At 1.5% strain, the total dislocation density is about 8 $\mu m^{-2}$, about 5 times the initial dislocation density. There are several mechanisms accounting for dislocation multiplication in our model. Forest dislocations on other slip systems can act as pinning points. Therefore, according to the transport equation, dislocations will bow out, leading to an increase in dislocation line lengths. When double cross slip happens, dislocation density will increase (Hirth and Lothe, 1982; Hull and Bacon, 2011). Our model is able to handle such processes naturally, since the cross slip terms connecting every two collinear slip system pair are included in equation (17). Glissile junctions are another important source of dislocation multiplication, as the junction segment can bow out on its slip plane (Stricker et al., 2018; Stricker and Weygand, 2015) and ultimately detach. As depicted in Figure 12(b), the dislocation densities on the individual slip systems appear as two groups. The densities on the 8 active slip systems increase more or less similarly, with some fluctuations, while the dislocation densities on the 4 inactive slip systems increases at a significantly lower rate. From a strict crystal plasticity point of view, the density on the inactive systems should not increase at all. However, glissile junctions couple the "inactive" slip systems with the active ones, and cross slip also leads to density evolution. For example, slip system $(111)[01\bar{1}]$ (numbered as slip system 1) and slip system $(\bar{1}11)[101]$ (numbered as slip system 3) are active slip systems and they can form a glissile junction on slip system $(\bar{1}11)[110]$ (numbered as slip system 11), which increases its dislocation density. Another reason for the finite nonzero increase in dislocation density on the inactive slip systems is that, although the Schmid factor is zero on theses slip systems, the local resolved shear stress is not usually precisely zero everywhere, since it includes contributions not only from the external load, but also from the stress field arising from dislocations. This fact enables even those dislocations on the nominally inactive slip systems to move.

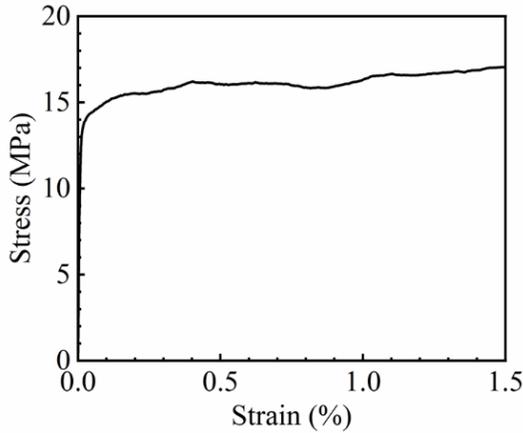

Figure 11 Stress-strain curve.

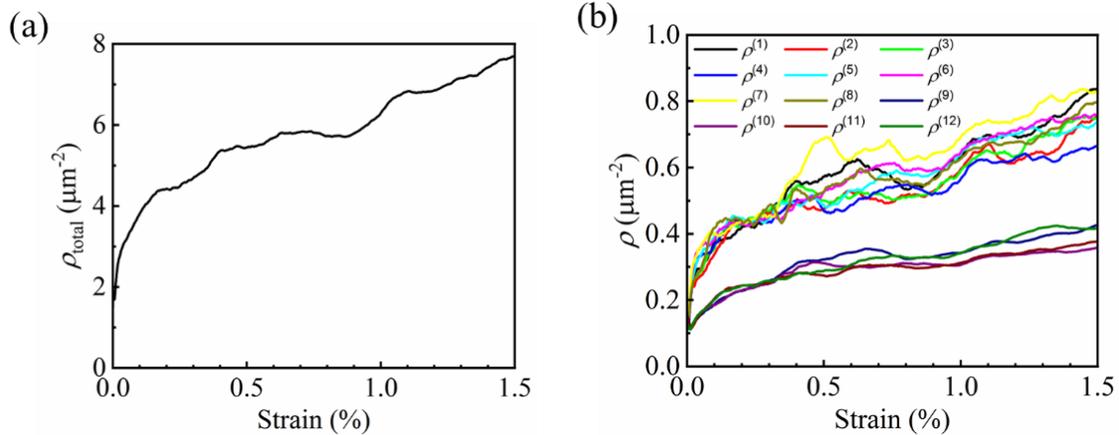

Figure 12 Evolution of the dislocation density. (a) Total dislocation density and (b) dislocation densities on individual slip systems.

Figure 13 shows the dislocation microstructure evolution on a (111) slip plane. The location of the section is shown in Figure 13(a). Figure 13(b), (c) and (d) show the total dislocation density at different strains. The total dislocation density is defined as the sum of dislocation densities on all slip systems, $\rho_{\text{total}} = \sum_k \|\boldsymbol{\rho}^{(k)}\|$, where $\boldsymbol{\rho}^{(k)}$ is the dislocation density vector on slip system $k$. The arcs in Figure 13(b) are the dislocation loops on that slip system, while the diffuse dots are dislocation loops on the other slip systems piercing the plane and serving as forest dislocations.

As the strain increases, dislocation density propagates over the domain through dislocation transport and dislocation reactions. Dislocations accumulate at some locations, forming a heterogeneous pattern. Figure 13(d) shows that dislocations are likely to accumulate in a pattern along three specific directions. One preferred direction is horizontal and the other two are oriented $\pm\pi/3$ from the horizontal line. These directions are the intersections of the slip plane with the other three slip planes of the FCC crystal. These dislocation walls are not stationary. They evolve as the strain increases. The dislocation walls are at different locations at different strains. Figure 14 shows another cross section of the dislocation microstructure pattern. As it is not a slip plane, there are no arcs in the initial configuration and all dislocation loops pierce this plane, see Figure 14(b). The dislocation pattern in Figure 14 is different from that in Figure 13. Dislocation walls are formed at angles of $\pm\pi/4$ from the horizontal line. They are shorter and less prominent than those in Figure 13.

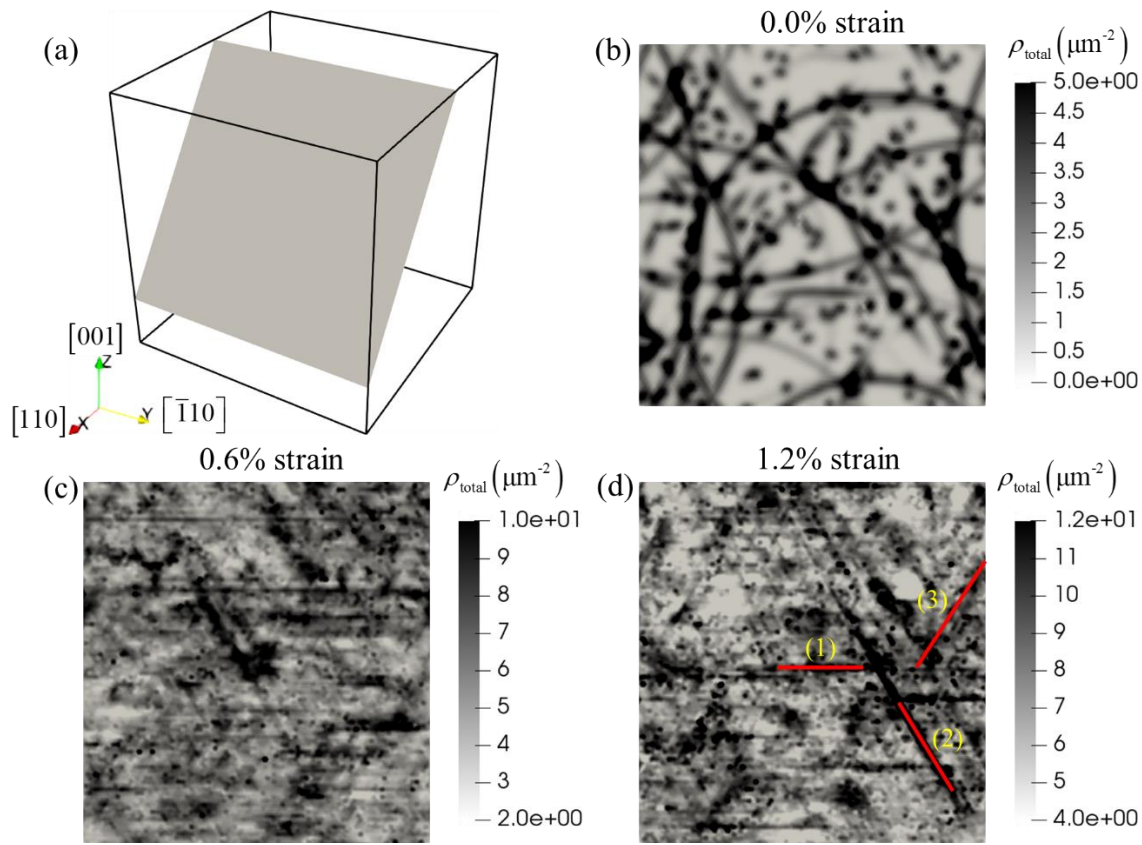

Figure 13 Dislocation microstructure evolution. (a) The location of the plane of view in the domain, which coincides with the (111) slip plane. (b) Initial total dislocation density. (c)

Page | 27

Total dislocation density at 0.6% strain. (d) Total dislocation density at 1.2% strain. The red lines show three directions along which dislocations accumulate the most.

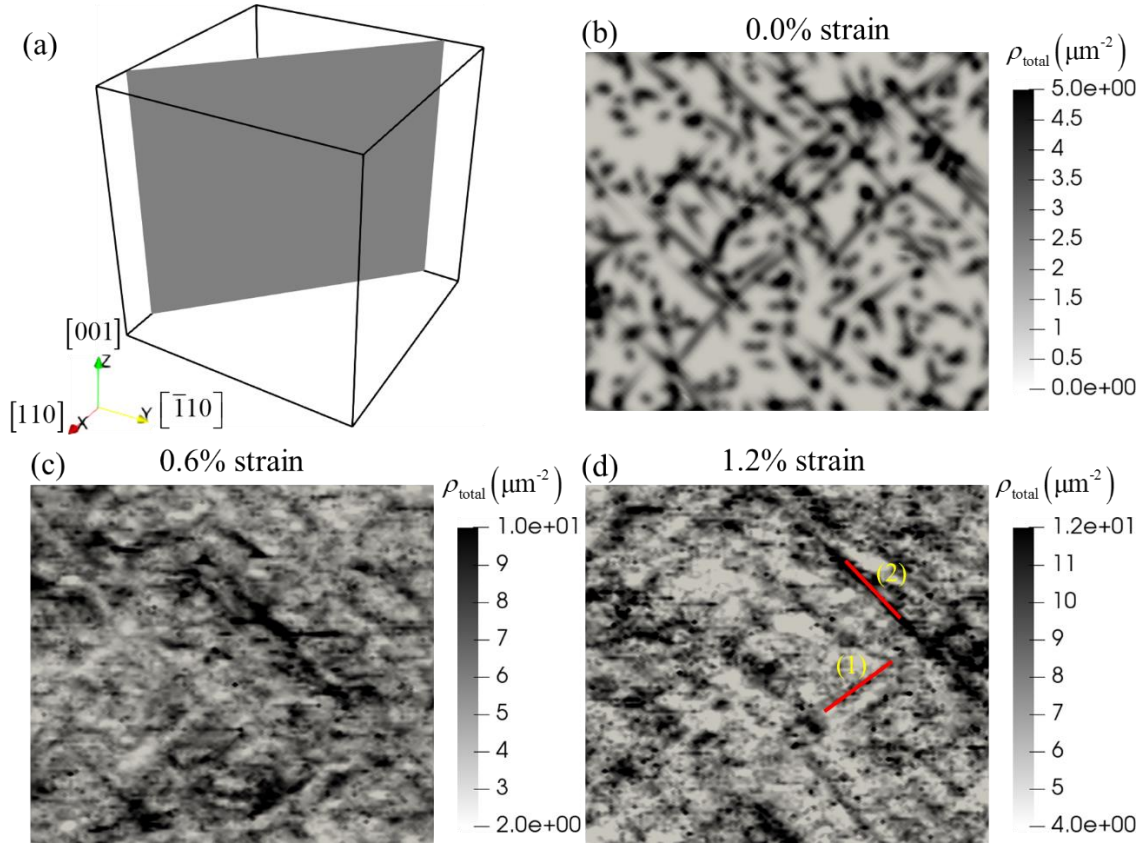

Figure 14 Dislocation microstructure evolution. (a) The location of the plane of view in the domain. (b) Initial total dislocation density. (c) Total dislocation density at 0.6% strain. (d) Total dislocation density at 1.2% strain. The red lines show three directions along which dislocations accumulate the most.

# 6 Concluding remarks

Two algorithms for updating the plastic distortion tensor and internal stress tensor were presented and tested in this paper. The first algorithm performs a direct time integration of the plastic distortion rate computed from Orowan's law, and the second utilizes the field dislocation mechanics formulation (Acharya and Roy, 2006; Brenner et al., 2014; Roy and Acharya, 2006, 2005). These two algorithms were introduced along with a full description of the vector-density based continuum dislocation dynamics approach, consisting of two sets of equations. One set of equations describes the crystal mechanics of the problem, with the kinematics and stress equilibrium parts, and the other set consists of the transport-reaction equations of continuum dislocation dynamics. The update algorithm based on field dislocation mechanics calculates the incompatible part of the plastic distortion directly from the dislocation density field, regardless of the dislocation evolution history.

While direct time integration of the plastic distortion rate accumulates numerical errors over time, the algorithm based on field dislocation mechanics was found to predict a plastic distortion field that is consistent with the instantaneous dislocation density field irrespective of time. However, this gain in accuracy incurs a computational cost. The use of field dislocation mechanics requires that one solves two more sets of equations for the compatible and incompatible parts of the plastic distortion. This method was explained. It consists of a staggered numerical scheme consisting of alternating steps of solving the stress equilibrium problem, followed by an update of the dislocation density effected by solving the dislocation transport-reaction equations. This sequence then is repeated. In this scheme, the first step yields the resolved shear stress required to compute the dislocation velocity, while the second step helps update the plastic strain required to solve the stress equilibrium problem. The stress field was solved by a standard Galerkin finite element method, and the dislocation dynamics calculations are performed using a least squares finite element method.

A simple test example was performed that clearly shows that the update algorithm based on field dislocation mechanics predicts a more accurate plastic distortion distribution and stress field than the older time integration method. The relative influences of the accuracy of the compatible and incompatible parts of the plastic distortion on stress field fidelity also was discussed. The conclusion is that the errors developed in the stress field mainly derive from the manner in which the incompatible part of the plastic distortion is calculated, and these errors can be reduced using the improved update algorithm presented herein. This algorithm was used to simulate the behavior of FCC steel under uniaxial loading and multiple slip conditions. The expected trends for the mechanical behavior of a FCC metal at small strain were obtained. All known major dislocation multiplication mechanisms were included in the model.

The evolution of the dislocation microstructures also was illustrated. In the continuum dislocation dynamics framework presented here, more mechanisms regarding dislocation reactions can be included to study the mechanical behavior of the material at microscale. For example, more accurate dislocation junction reaction rates can be included from DDD, and interactions of dislocation with point defects can be introduced.

**Acknowledgements**

The authors are grateful for the support from the Naval Nuclear Laboratory, operated by Fluor Marine Propulsion, LLC for the US Naval Reactors Program. A. El-Azab assisted with the formulation of field dislocation mechanics with support from the US Department of Energy, Office of Science, Division of Materials Sciences and Engineering, through award number DE-SC0017718. K. Starkey assisted with the deployment of the numerical approach and was supported by the National Science Foundation, Division of Civil, Mechanical, and Manufacturing Innovation (CMMI), through award number 1663311 at Purdue University.

**Appendix: Finite element formulations of the equations**

The Galerkin finite element method is commonly used to solve mechanical problems (Belytschko et al., 2013). By taking the weak form of equation (11) and discretizing it, a standard linear algebraic system is obtained for the nodal displacements,

$$[K_u]\{u\} = \{P_u\}, \tag{A1}$$

where $\{u\}^T = \{u_{11}, u_{21}, u_{31}, ..., u_{1I}, u_{2I}, u_{3I}\}^T$, with $I$ being the number of nodes per element, is the collection of the nodal values of displacement in an element, and $[K_u]$ is the element stiffness matrix expressed in the form

$$[K_u] = \int_\Omega [B]^T [C][B] dV, \tag{A2}$$

with $[C]$ being the elastic stiffness matrix. $[B]$ contains the derivatives of the shape functions of the element. $\{P_u\}$ is the load vector, which is contributed in the current case by the plastic distortion term in the stress equilibrium equation (11). Considering the two update algorithms explained in section 3, this load vector has the form:

$$\begin{aligned} \{P_u\} &= \int_\Omega [B]^T [C][E][N_\gamma]\{\gamma\} dV & \text{with algorithm 1,} \\ \{P_u\} &= \int_\Omega [B]^T [C]\big([B]\{z\} - [N_\chi]\{\chi\}\big) dV & \text{with algorithm 2,} \end{aligned} \tag{A3}$$

where $\{\gamma\}$, $\{z\}$ and $\{\chi\}$ are the corresponding node values of $\gamma$, $\mathbf{z}$ and $\boldsymbol{\chi}$. $[N_\gamma]$ and $[N_\chi]$ are the matrices that contain shape functions of the element. $[E]$ defines the slip normal and slip direction of each slip system.

Equation (A1) is the linear algebraic system to be solved for displacement. The stiffness matrix $[K_u]$ of the discretized system remains unchanged at different time steps, so it is calculated and assembled only at the initial time step. The load vector $\{P_u\}$ changes due to the evolution of $\{\gamma\}$ or $\{z\}$ and $\{\chi\}$, and it is updated at every step.

Next, we turn attention to the dislocation transport problem. We use a 2D representation of the 3D dislocation vector field $\boldsymbol{\rho}^\alpha$ to reduce the number of degrees of freedom by considering a set of local coordinate systems with base vectors $(\hat{\mathbf{e}}_1^\alpha, \hat{\mathbf{e}}_2^\alpha, \hat{\mathbf{e}}_3^\alpha)$ defined such that, for each slip system, $\hat{\mathbf{e}}_1^\alpha$ is in the direction of the Burgers vector, $\hat{\mathbf{e}}_3^\alpha$ is in the direction of the normal to the slip plane, and $\hat{\mathbf{e}}_2^\alpha = \hat{\mathbf{e}}_3^\alpha \times \hat{\mathbf{e}}_1^\alpha$. With these relationships in hand, the vector and tensor quantities can be written in both global $(\mathbf{e}_1, \mathbf{e}_2, \mathbf{e}_3)$ and local $(\hat{\mathbf{e}}_1, \hat{\mathbf{e}}_2, \hat{\mathbf{e}}_3)$ coordinates. For example,

$$\boldsymbol{\rho}^\alpha = \rho_i^\alpha \mathbf{e}_i = \hat{\rho}_j^\alpha \hat{\mathbf{e}}_j^\alpha \quad \text{and} \quad \boldsymbol{\sigma} = \sigma_{ij} \mathbf{e}_i \mathbf{e}_j = \hat{\sigma}_{kl} \hat{\mathbf{e}}_k^\alpha \hat{\mathbf{e}}_l^\alpha, \tag{A4}$$

with the components transforming using the matrix $Q_{ij} = \mathbf{e}_i \cdot \hat{\mathbf{e}}_j^\alpha$ according to

$$\hat{\rho}_i^\alpha = Q_{ji}^\alpha \rho_j^\alpha \quad \text{and} \quad \hat{\sigma}_{ij} = Q_{ki}^\alpha Q_{lj}^\alpha \sigma_{kl}. \tag{A5}$$

In the local coordinate systems, however, the dislocation density and velocity on a given slip system have only two components,

$$\boldsymbol{\rho} = \hat{\rho}_1 \hat{\mathbf{e}}_1 + \hat{\rho}_2 \hat{\mathbf{e}}_2 \quad \text{and} \quad \mathbf{v} = \hat{v}_1 \hat{\mathbf{e}}_1 + \hat{v}_2 \hat{\mathbf{e}}_2, \tag{A6}$$

so that the dislocation density evolution equation (6) can be written in its component form

$$\dot{\hat{\rho}}_1^\alpha = \frac{\partial}{\partial \hat{x}_2}(\hat{v}_1^\alpha \hat{\rho}_2^\alpha - \hat{v}_2^\alpha \hat{\rho}_1^\alpha) \quad \text{and} \quad \dot{\hat{\rho}}_2^\alpha = -\frac{\partial}{\partial \hat{x}_1}(\hat{v}_1^\alpha \hat{\rho}_2^\alpha - \hat{v}_2^\alpha \hat{\rho}_1^\alpha). \tag{A7}$$

The last equations can be rewritten in a matrix form,

$$[A_t] \begin{Bmatrix} \dot{\hat{\rho}}_1^\alpha \\ \dot{\hat{\rho}}_2^\alpha \end{Bmatrix} = \left\{ [A_0] + [A_1] \frac{\partial}{\partial \hat{x}_1} + [A_2] \frac{\partial}{\partial \hat{x}_2} \right\} \begin{Bmatrix} \hat{\rho}_1^\alpha \\ \hat{\rho}_2^\alpha \end{Bmatrix}, \tag{A8}$$

where

$$[A_t] = \begin{bmatrix} 1 & 0 \\ 0 & 1 \end{bmatrix}, \quad [A_0] = \begin{bmatrix} -\dfrac{\partial \hat{v}_2^\alpha}{\partial \hat{x}_2} & \dfrac{\partial \hat{v}_1^\alpha}{\partial \hat{x}_2} \\ \dfrac{\partial \hat{v}_2^\alpha}{\partial \hat{x}_1} & -\dfrac{\partial \hat{v}_1^\alpha}{\partial \hat{x}_1} \end{bmatrix}, \quad [A_1] = \begin{bmatrix} 0 & 0 \\ \hat{v}_2^\alpha & -\hat{v}_1^\alpha \end{bmatrix}, \quad [A_2] = \begin{bmatrix} -\hat{v}_1^\alpha & \hat{v}_1^\alpha \\ 0 & 0 \end{bmatrix}. \quad (A9)$$

Equation (A8) is discretized in time by a first order implicit Euler method, giving

$$\left\{ -[A_t] + \Delta t [A_0] + \Delta t [A_1] \dfrac{\partial}{\partial \hat{x}_1} + \Delta t [A_2] \dfrac{\partial}{\partial \hat{x}_2} \right\} \begin{Bmatrix} \hat{\rho}_1^\alpha \\ \hat{\rho}_2^\alpha \end{Bmatrix}^{n+1} = -[A_t] \begin{Bmatrix} \hat{\rho}_1^\alpha \\ \hat{\rho}_2^\alpha \end{Bmatrix}^n, \quad (A10)$$

with $\{\cdot\}^{n+1}$ and $\{\cdot\}^n$ representing the unknown variable at time steps $n+1$ and $n$, respectively. The velocity in $A_0$, $A_1$ and $A_2$ is taken at time step $n$. If we write the unknown variable in terms of the nodal values $\hat{\rho}_i^I$ with the aid of shape functions of a finite element, which denote the $i$th component of $I$th node,

$$\begin{Bmatrix} \hat{\rho}_1^\alpha \\ \hat{\rho}_2^\alpha \end{Bmatrix} = [N_\rho] \begin{Bmatrix} \hat{\rho}_1^1 \\ \hat{\rho}_2^1 \\ \hat{\rho}_1^2 \\ \hat{\rho}_2^2 \\ \vdots \\ \hat{\rho}_1^M \\ \hat{\rho}_2^M \end{Bmatrix}, \quad \dfrac{\partial}{\partial \hat{x}_1} \begin{Bmatrix} \hat{\rho}_1^\alpha \\ \hat{\rho}_2^\alpha \end{Bmatrix} = [B_1] \begin{Bmatrix} \hat{\rho}_1^1 \\ \hat{\rho}_2^1 \\ \hat{\rho}_1^2 \\ \hat{\rho}_2^2 \\ \vdots \\ \hat{\rho}_1^M \\ \hat{\rho}_2^M \end{Bmatrix}, \quad \dfrac{\partial}{\partial \hat{x}_2} \begin{Bmatrix} \hat{\rho}_1^\alpha \\ \hat{\rho}_2^\alpha \end{Bmatrix} = [B_2] \begin{Bmatrix} \hat{\rho}_1^1 \\ \hat{\rho}_2^1 \\ \hat{\rho}_1^2 \\ \hat{\rho}_2^2 \\ \vdots \\ \hat{\rho}_1^M \\ \hat{\rho}_2^M \end{Bmatrix} \quad (A11)$$

where $M$ is the number of nodes of the element and

$$[N_\rho] = \begin{bmatrix} N_1 & 0 & N_2 & 0 & \cdots & N_M & 0 \\ 0 & N_1 & 0 & N_2 & \cdots & 0 & N_M \end{bmatrix}$$

$$[B_1] = \begin{bmatrix} \dfrac{\partial N_1}{\partial \hat{x}_1} & 0 & \dfrac{\partial N_2}{\partial \hat{x}_1} & 0 & \cdots & \dfrac{\partial N_M}{\partial \hat{x}_1} & 0 \\ 0 & \dfrac{\partial N_1}{\partial \hat{x}_1} & 0 & \dfrac{\partial N_2}{\partial \hat{x}_1} & \cdots & 0 & \dfrac{\partial N_M}{\partial \hat{x}_1} \end{bmatrix} \quad (A12)$$

$$[B_2] = \begin{bmatrix} \dfrac{\partial N_1}{\partial \hat{x}_2} & 0 & \dfrac{\partial N_2}{\partial \hat{x}_2} & 0 & \cdots & \dfrac{\partial N_M}{\partial \hat{x}_2} & 0 \\ 0 & \dfrac{\partial N_1}{\partial \hat{x}_2} & 0 & \dfrac{\partial N_2}{\partial \hat{x}_2} & \cdots & 0 & \dfrac{\partial N_M}{\partial \hat{x}_2} \end{bmatrix}$$

For ease of reading, the superscript $\alpha$ has been dropped in the nodal quantities. Eq. (A10) can be rewritten for the node values as follows:

$$[L_\rho]\{\hat{\rho}\}^{n+1} = [W]\{\hat{\rho}\}^n, \tag{A13}$$

with

$$[L_\rho] = -[A_t][N_\rho] + \Delta t[A_0][N_\rho] + \Delta t[A_1][B_1] + \Delta t[A_2][B_2],$$
$$[W] = -[A_t][N_\rho], \tag{A14}$$
$$\{\hat{\rho}\} = \left\{\hat{\rho}_1^1, \hat{\rho}_2^1, \hat{\rho}_1^2, \hat{\rho}_2^2, \cdots, \hat{\rho}_1^I, \hat{\rho}_2^I\right\}^T.$$

The system of transport equations for dislocation density evolution is hyperbolic. Some level of numerical diffusion and dispersion is always anticipated with the discretization of this kind of partial differential equations (Varadhan et al., 2006). A least squares finite element method (LSFEM) (Jiang, 2013) is used in our model to stabilize dispersion. The least squares residual corresponding to equation (A13) is given by

$$R = \frac{1}{2}\int_\Omega \left([L_\rho]\{\hat{\rho}\}^{n+1} - [W]\{\hat{\rho}\}^n\right)^T \left([L_\rho]\{\hat{\rho}\}^{n+1} - [W]\{\hat{\rho}\}^n\right) d\Omega. \tag{A15}$$

A key attribute of dislocations is that they do not end inside a crystal. In continuum dislocation theory, this requirement is enforced by imposing the condition that the dislocation tensor $\boldsymbol{\alpha}$ is divergence free. That is,

$$\nabla \cdot \boldsymbol{\alpha} = \mathbf{0}. \tag{A16}$$

Theoretically, equation (A16) is satisfied in our model because of equation (2). However, in the numerical implementation, a divergence-free condition must be used as a constraint on the evolution of the field $\boldsymbol{\rho}$ to ensure numerical accuracy. An additional term representing this constraint is thus added to the residual in equation (A15), which has the form

$$R' = \frac{1}{2}ch^2\int_\Omega |\nabla \cdot (\boldsymbol{\rho}^\alpha - \boldsymbol{\rho}_{cp}^\alpha)|^2\, d\Omega, \tag{A17}$$

where $c$ is a control parameter, $h$ is the mesh size, and $\boldsymbol{\rho}_{cp}^\alpha$ is the coupling term due to dislocation reactions and cross slip, as defined in Eq. (17). Including $\boldsymbol{\rho}_{cp}^\alpha$ in Eq. (A17) is based on the fact that when dislocation reaction happens, the dislocation line can not be considered as continuous and connected on a individual slip system anymore unless we take the reaction portion back. In a matrix form, the last residual becomes

$$R' = \frac{1}{2}ch^2\int_\Omega \left([B_d]\{\hat{\rho} - \hat{\rho}_{cp}\}^{n+1}\right)^2 d\Omega, \tag{A18}$$

where

$$[B_d] = \begin{bmatrix} \dfrac{\partial N_1}{\partial x_1} & \dfrac{\partial N_1}{\partial x_2} & \dfrac{\partial N_2}{\partial x_1} & \dfrac{\partial N_2}{\partial x_2} & \cdots & \dfrac{\partial N_I}{\partial x_1} & \dfrac{\partial N_I}{\partial x_2} \end{bmatrix}. \quad (A19)$$

Adding equation (A18) into equation (A15) and taking the first variation of the function to be zero, we have

$$\left(\delta\{\hat{\rho}\}^{n+1}\right)^T \int_\Omega [L_\rho]^T \left([L_\rho]\{\hat{\rho}\}^{n+1} - [W]\{\hat{\rho}\}^n\right) + ch^2 [B_d]^T [B_d]\{\hat{\rho} - \hat{\rho}_{\text{cp}}\}^{n+1} \, d\Omega = 0. \quad (A20)$$

As $\delta\{\rho\}^{n+1}$ is arbitrary, Eq. (A20) can be rewritten into the form

$$\begin{aligned}
&\int_\Omega \left([L_\rho]^T [L_\rho] + ch^2 [B_d]^T [B_d]\right) d\Omega \{\hat{\rho}\}^{n+1} \\
&= \int_\Omega \left([L_\rho]^T [W]\{\hat{\rho}\}^n + ch^2 [B_d]^T [B_d]\{\hat{\rho}_{\text{cp}}\}^{n+1}\right) d\Omega
\end{aligned}, \quad (A21)$$

which yields the algebraic system

$$[K_\rho]\{\hat{\rho}\}^{n+1} = \{P_\rho\}, \quad (A22)$$

with

$$\begin{aligned}
[K_\rho] &= \int_\Omega \left([L_\rho]^T [L_\rho] + ch^2 [B_d]^T [B_d]\right) d\Omega, \\
\{P_\rho\} &= \int_\Omega \left([L_\rho]^T [W]\{\hat{\rho}\}^n + ch^2 [B_d]^T [B_d]\{\hat{\rho}_{\text{cp}}\}^{n+1}\right) d\Omega
\end{aligned}. \quad (A23)$$

The solution of Eq. (A22) yields the dislocation density evolution. The stiffness matrix $[K_\rho]$ and the load vector $\{P_\rho\}$ depend on the problem information at time step $n$ and are updated at every time step.

When we use field dislocation mechanics to calculate the incompatible and compatible parts of the plastic distortion, Eqs. (22) and (23) are also solved by finite element methods. For the incompatible part of the plastic distortion, Eq. (22) can be written in a matrix form,

$$\begin{bmatrix} 0 & -\dfrac{\partial}{\partial x_3} & \dfrac{\partial}{\partial x_2} \\ \dfrac{\partial}{\partial x_3} & 0 & -\dfrac{\partial}{\partial x_1} \\ -\dfrac{\partial}{\partial x_2} & \dfrac{\partial}{\partial x_1} & 0 \\ \dfrac{\partial}{\partial x_1} & \dfrac{\partial}{\partial x_2} & \dfrac{\partial}{\partial x_3} \end{bmatrix} \begin{bmatrix} \chi_{11} & \chi_{12} & \chi_{13} \\ \chi_{21} & \chi_{22} & \chi_{23} \\ \chi_{31} & \chi_{32} & \chi_{33} \end{bmatrix} = \begin{bmatrix} \alpha_{11} & \alpha_{12} & \alpha_{13} \\ \alpha_{21} & \alpha_{22} & \alpha_{23} \\ \alpha_{31} & \alpha_{32} & \alpha_{33} \\ 0 & 0 & 0 \end{bmatrix}. \tag{A24}$$

Using $[L_\chi]$ to represent the linear operator, the equation can be expressed in the form,

$$[L_\chi]\{\chi\} = \{\alpha\}. \tag{A25}$$

By using the least squares finite element method to solve this partial differential equation, we finally reach an algebraic system of the form,

$$[K_\chi]\{\chi\} = \{P_\chi\}, \tag{A26}$$

where the stiffness matrix $[K_\chi]$ and load vector $\{P_\chi\}$ are given by

$$\begin{aligned} [K_\chi] &= \int_\Omega [L]^{\mathrm{T}}[L]\,\mathrm{d}\Omega \\ \{P_\chi\} &= \int_\Omega [L]^{\mathrm{T}}\{\alpha\}\,\mathrm{d}\Omega \end{aligned}. \tag{A27}$$

For the compatible part of the plastic distortion, Eq. (23) can be rewritten for a time step $\Delta t$ in the form

$$\nabla \cdot \left( \nabla \mathbf{z}^{(n+1)} - \nabla \mathbf{z}^{(n)} + \Delta t \sum \left( \mathbf{v}^{(n)} \times \boldsymbol{\alpha}^{(n+1)} \right) \right) = 0. \tag{A28}$$

A standard Galerkin finite element method is used to solve this partial differential equation, the weak form of which is

$$\int_\Omega \left( \nabla \mathbf{z}^{(n+1)} - \nabla \mathbf{z}^{(n)} + \Delta t \sum \left( \mathbf{v}^{(n)} \times \boldsymbol{\alpha}^{(n+1)} \right) \right) \cdot \nabla \delta \mathbf{z}^{(n+1)} \mathrm{d}\Omega = 0. \tag{A29}$$

The corresponding algebraic system has the form

$$[K_z]\{z\}^{n+1} = \{P_z\}, \tag{A30}$$

where the stiffness matrix $[K_z]$ and load vector $\{P_z\}$ are given by

$$[K_z] = \int_\Omega [B]^T [B] \, d\Omega$$
$$\{P_z\} = \int_\Omega [B]^T \left([B]\{z\}^n + \{p\}\right) d\Omega \quad .\tag{A31}$$

The $[B]$ matrix is the same as in equation (A2) and $\{p\}$ denotes $\Delta t \sum \left(\mathbf{v}^{(n)} \times \boldsymbol{\alpha}^{(n+1)}\right)$.